\documentclass[prl,reprint, longbibliography,superscriptaddress]{revtex4-2}

\usepackage{braket}
\usepackage{graphicx}
\usepackage{amsmath}
\usepackage{hyperref}
\usepackage[all]{hypcap}
\hypersetup{colorlinks=True}
\usepackage{color}
\usepackage{subfigure}
\usepackage{placeins}
\usepackage{times}
\usepackage{bbold}
\DeclareMathOperator{\Tr}{Tr}

\begin{document}
\author{Xin H. H. Zhang}
\email[Corresponding author: ]{physicsxinzhang@gmail.com}
\affiliation{Dahlem  Center  for  Complex  Quantum  Systems  and  Fachbereich  Physik, Freie  Universit\"at  Berlin,  14195  Berlin,  Germany}

\author{S. H. L. Klapp}
\affiliation{Institute for Theoretical Physics, 
Technische Universität Berlin, 10623 Berlin, Germany}

\author{A. Metelmann}
\email[ Corresponding author: ]{anja.metelmann@kit.edu}
\affiliation{Dahlem  Center  for  Complex  Quantum  Systems  and  Fachbereich  Physik, Freie  Universit\"at  Berlin,  14195  Berlin,  Germany}
\affiliation{Institute for Theory of Condensed Matter, Karlsruhe Institute of Technology, 76131 Karlsruhe, Germany}
\affiliation{Institute for Quantum Materials and Technology, Karlsruhe Institute of Technology, 76344 Eggenstein-Leopoldshafen, Germany}

\date{April 5, 2022}

\title{Embedding of Time-Delayed Quantum Feedback in a Nonreciprocal Array}

\begin{abstract}

Time-delayed quantum feedback is a fast and efficient method to control and stabilize few and many-body quantum systems. However, a proper understanding of such systems stays opaque due to the non-Markovian nature of the feedback protocol. Here, we present a method of encoding time-delayed quantum feedback into a chain of nonreciprocally coupled auxiliary oscillators. Our approach serves as a novel method of introducing time-delayed quantum feedback and provides the advantage of large tunability. Importantly, within our approach the original non-Markovian system is embedded into an enlarged Markovian open quantum system, which can be treated efficiently with quantum many-body methods.  As an exemplary illustration, we apply our Markovian embedding approach to the paradigmatic case of a driven-dissipative atom in front of a mirror.

\end{abstract}

\maketitle

%%%%%%%%%%%%%%%%%%%%%%%%%%%%%%%%%%%%%%%%%%%%%%%%
%Introduction

\begin{figure}[bt]
	\center
	\includegraphics[width=0.45\textwidth]{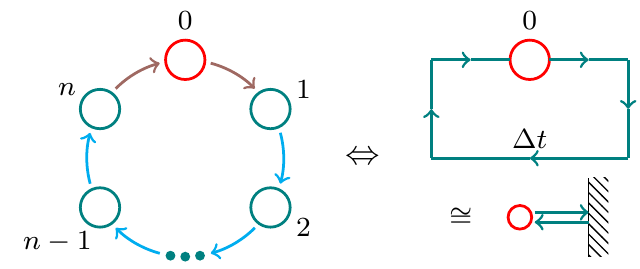}
	\put(-225,80){(a)}
    \put(-95,80){(b)}
    \put(-95,16){(c)}
	\caption{Schematic of the setup and the Markovian embedding approach. (a) The total system under study, which is composed of a quantum system of interest at site $0$ and $n$ auxiliary dissipative oscillators. The couplings (in Hamiltonian and jump operators) between them are engineered such that the dynamics is chiral. (b) After integrating out the $n$ auxiliary oscillators, we obtain a time-delayed feedback for site $0$ with delay time $\Delta t$. It turns out that this time-delayed dynamics is equivalent to that of a quantum system in front of a mirror [panel (c)], when normally-ordered operators are considered.  Since the total system shown in (a) is a  Markovian open quantum system, we therefore provide a Markovian embedding of the time-delayed dynamics in (b) and (c), by encoding memory into the chain of auxiliary oscillators.
	}
	\label{setup}
\end{figure}

With the rapid advances of various quantum platforms   \cite{HarocheBook2006,LaddNature2010,BlochNatPhys2012,ClerkNatPhy2020,BlaisRMP2021}, time-delayed quantum feedback has been an increasingly important element in designing and controlling quantum devices \cite{WisemanBook2014,PyragasPLA1992,CarmelePRL2013,HeinPRL2014,DroennerPRA2019}, 
as well as for processing of quantum information in quantum networks
\cite{ZhengPRL2013,PichlerPNAS2017,KockumPRL2018}, especially when the time delay between quantum nodes becomes significant \cite{KimbleNature2008}. A paradigmatic example for a quantum system with time-delayed feedback is an atom in front of a mirror [Fig.\,\ref{setup}(c)] or similarly an atom coupled with a semi-infinite/chiral waveguide (see recent experiments \cite{vanLooScience2013,HoiNatPhys2015,ManentiNatCom2017,KannanNature2020}). The atom at time $t$ undergoes here a self-driven cycle of absorbing excitations emitted at an earlier time $t-\Delta t$, 
which turns the dynamics of the system non-Markovian. Such memory effects are a crucial characteristic of the quantum dynamics with time-delayed feedback \cite{BreuerRMP2016,deVegaRMP2017}, and are the origin of the underdeveloped understanding of the effects of this sort of autonomous feedback protocols. Current theoretical approaches to time-delayed quantum feedback (e.g.\,\cite{DornerPRA2002,CarmelePRL2013,TufarelliPRA2013,LaaksoPRL2014,DiazCamachoPRA2014,FangPRA2015,GrimsmoPRL2015,PichlerPRL2016,GuoPRA2017,CalajoPRL2019,SinhaPRL2020,CrowderPRA2020,ArranzRegidorPRR2021}) usually become intractable beyond the few-photon regime or the transient regime. To explore previously inaccessible regimes, novel approaches are needed.      

In this letter, we introduce a novel method for tackling time-delayed quantum feedback, which embeds the time-delayed memory into a chain of dissipative oscillators whose dynamics is nonreciprocal [Fig.\,\ref{setup}(a)]. This approach is inspired by the Markovian embedding approach to time-delayed Langevin equations (see e.g.\,\cite{ZwanzigJSP1973,SieglePRE2010,LoosJSP2019,LoosNJP2020} and a brief introduction in \cite{SupMat}), which embeds delayed feedback into auxiliary variables based on the linear chain trick \cite{CushingBook1977, SmithBook2010}. 
Here, in the quantum setting, we show that the resulting system is a Markovian open quantum system, which can be analyzed efficiently using quantum many-body methods. 
The purpose here is twofold:   (i) our Markovian embedding approach serves as an efficient method for analyzing and understanding time-delayed quantum feedback. For illustrative purpose, we apply this to the paradigmatic atom-mirror case and characterize its long-time dynamics using exact diagonalization (in the linear case) and matrix product state methods (in the nonlinear case). (ii) the proposed setup represents a novel way of introducing time-delayed feedback, which features large tunability on the memory kernel by tuning parameters of the auxiliary oscillators. It can therefore enable us to explore more feedback-induced effects, considering its solvability.

{\it The model and its dynamics.---}To introduce time-delayed feedback on a system of interest at site $0$, we couple it with the two ends of a chain of $n$ auxiliary dissipative oscillators as shown in Fig.\,\ref{setup}. The key is that the dynamics in the chain is engineered such that signals can only propagate unidirectionally from the oscillator at site $0$ to the oscillator at site $n$. Thus, the system at site $0$ at time $t$ is affected by its  historical quantum state at time $t-\Delta t$, where $\Delta t$ the time of propagation from site $0$ to site $n$. In this letter, we show that this intuition indeed works and the desired time-delayed quantum feedback is realized.  

As noticed, the crucial ingredient here is that the oscillators are coupled nonreciprocally, which means that the dynamics of site $j$ depends on site $j-1$ but {\it not} vice versa, which can be realized by reservoir engineering using the recipe developed in Ref.\,\cite{MetelmannPRX2015,MetelmannPRA2017}. 
For example, to engineer a nonreciprocal coupling  of two oscillators with annihilation operators $a_{1,2}$, the coherent hopping interaction
$H_{12} = \frac{\gamma}{2} (a_{1} a_{2}^{\dagger} + \text{h.c.}) $
is combined with the corresponding dissipative hopping process, modeled via the jump operator $J_{12} = \sqrt{\gamma} (a_{1} - i a_{2})$, in a Lindblad master equation
\begin{equation}
    \frac{d}{dt} \rho_{12} = -i [H_{12}, \rho_{12}] + J_{12} \rho_{12}  J_{12}^{\dagger} - \frac{1}{2} \{ \rho_{12} ,  J_{12}^{\dagger} J_{12} \},
\end{equation}
which coincides with the master equation for cascaded quantum systems \cite{GardinerZollerBook}.
The corresponding Heisenberg-Langevin equations yield (for details see \cite{SupMat})
\begin{equation}\label{2OscillatorHLE}  
    \begin{split}
        \frac{d}{dt} a_{1} &= - \frac{\gamma}{2} a_{1} + \eta(t),\\
        \frac{d}{dt} a_{2} &=- \frac{\gamma}{2} a_{2} -i\gamma a_{1} + i\eta(t),
    \end{split}
\end{equation}
where $\eta(t) = - \sqrt{\gamma/2\pi} \int dk\,b_{k}(0)e^{-ikt} $ is a Gaussian white noise with $\braket{\eta(t)\eta^{\dagger}(t^{\prime})} = \gamma \delta(t-t^{\prime})$, $\braket{\eta^{\dagger}(t) \eta(t^{\prime})} = 0$, and $\braket{\eta(t)\eta(t^{\prime})} = 0$. $b_{k}(0)$ is the bosonic annihilation operator at time $t=0$ for the environmental degrees of freedom of the two dissipative oscillators. Eq.\,\eqref{2OscillatorHLE} clearly shows that the dynamics is nonreciprocal since $a_{2}$ depends on $a_{1}$ but not vice versa. Generally, there is freedom in engineering $H_{12}$ and $J_{12}$ via a complex weight $\chi \in \mathbb{C}$ such that $H_{12} = \frac{\gamma}{2} (\chi^{*} a_{1} a_{2}^{\dagger} + \text{h.c.})$  and $J_{12} = \sqrt{\gamma} (a_{1} - i \chi a_{2})$, and this method is straightforwardly generalized to a network of oscillators \cite{MetelmannPRA2018}.

Applying the dissipation engineering protocol to the ring lattice as shown in Fig.\,\ref{setup}(a), we obtain clockwise nonreciprocal dynamics (site $0 \rightarrow 1 \rightarrow \ldots \rightarrow n \rightarrow 0$), captured in the Heisenberg-Langevin equations  (for details see \cite{SupMat}) 
\begin{equation}\label{ajEOM}
\begin{split}
    \frac{d}{dt} a_{0}=& -  \frac{1}{2}(\chi_{n,0}^2\gamma_{n,0}+\gamma_{0,1})  a_{0} -i\chi_{n,0} \gamma_{n,0} a_{n} + \xi_{0}(t) \\
    &+i [H_{0}, a_{0}], \\   
    \frac{d}{dt} a_{1}=& -  \frac{1}{2}(\chi_{0,1}^2 \gamma_{0,1} + \gamma)  a_{1} -i\chi_{0,1} \gamma_{0,1} a_{0} + \xi_{1}(t), \\ 
    \frac{d}{dt} a_{j}=& -  \gamma  a_{j} -i\gamma a_{j-1} + \xi_{j}(t),\,\,\text{for}\,\,2\leq j\leq (n-1), \\ 
    \frac{d}{dt} a_{n}=& -  \frac{1}{2}(\gamma+\gamma_{n,0})  a_{n} -i\gamma a_{n-1} + \xi_{n}(t), \\ 
\end{split}
\end{equation}
where we denote the decay rate and weight at bond $j\rightarrow j+1$ (with $n+1\equiv 0$) as $\gamma_{j,j+1}$ and $\chi_{j,j+1}$. 
Note that we account for  a local Hamiltonian $H_{0}$ at site $0$ which can be nonlinear. $\xi_{j}(t)$ are noise operators due to dissipation and we have chosen the weights $\chi$ to be real. Because each site $j$ has noises from two neighboring dissipators, $\xi_{j}(t)$ is given by a summation of two Gaussian white noises \begin{equation}\label{defXi}
    \xi_{j}(t) = - \sqrt{\frac{\gamma}{2\pi}} \int dk\,\left( i b_{k}^{(j-1,j)}(0) + b_{k}^{(j,j+1)}(0) \right)e^{-ikt},
\end{equation} 
where $b_{k}^{(j,j+1)}$ denotes bosonic operator for the bath of the bond $j\rightarrow j+1$. 

Like the case with two oscillators, the dynamics in this lattice is nonreciprocal, with site $j$ depends only on itself and site $j-1$ but not on site $j+1$. For simplification of analysis, in the current study, we let bonds $0\rightarrow 1$ and $n\rightarrow 0$ have free parameters while other bonds are homogeneous with decay rate $\gamma$ and weight $\chi=1$ [Fig.\,\ref{setup}(a)].  Though this choice is sufficient for the current study, in general, keeping more free parameters gives more tunability, which can be crucial in studies of other models.

{\it Time-delayed feedback mediated by the nonreciprocal chain.---}From Eq.\,\eqref{ajEOM}, we see that the dynamics for site $1,\ldots,n$ is linear, which means that we can analytically integrate out $a_{j>0}$ of the auxiliary oscillators. As shown in the following, with proper choice of parameters, we get $a_{n}(t) \propto a_{0} (t - \Delta t)$. After substituting it into the equation of motion (EOM) of $a_{0}$ in Eq.\,\eqref{ajEOM}, we obtain a time-delayed feedback for $a_{0}$ with discrete delay time $\Delta t$.

The linear EOM for $a_{j>0}$ in Eq.\,\eqref{ajEOM} can be solved with a Laplace transformation. In what follows we consider the case that all auxiliary oscillators are exposed to the same local damping $\gamma$, i.e., we set $\gamma_{n,0}= \gamma$ and $\gamma_{0,1} = \gamma/\chi_{0,1}^2$, for the general case please see \cite{SupMat}.
We obtain for the auxiliary oscillator operators in Laplace space
\begin{equation}\label{ajs}
    \tilde{a}_{j}(s) = \frac{-i\gamma_{j}}{ s +\gamma} \tilde{a}_{j-1}(s) + \frac{1}{ s +\gamma} \left(\tilde{\xi}_{j}(s) + a_j(0)\right),
\end{equation} 
with $\gamma_{1} = 1/\chi_{0,1}$ and $\gamma_{j}= \gamma$ for  $2 \leq j \leq n$.
Then iteratively, we get a relation between $\tilde{a}_{n}(s)$ and $\tilde{a}_{0}(s)$
\begin{equation}\label{anm1s}
\begin{split}
    \tilde{a}_{n}(s) =  & \frac{1}{\chi_{0,1}}\left( \frac{-i\gamma}{ s +\gamma} \right)^{n} \tilde{a}_{0}(s) %\\ &
    + \frac{i}{\gamma}\sum_{j=1}^{n} \left( \frac{-i\gamma}{s +\gamma} \right)^{n+1-j}   \tilde{\xi}^{\prime}_{j}(s),
\end{split}
\end{equation}
where we have defined $\tilde{\xi}^{\prime}_{j}(s) = \tilde{\xi}_{j}(s) + a_j(0)$.
Performing the inverse Laplace transformation of \eqref{anm1s}, we obtain
\begin{equation} \label{anm1t}
 a_{n}(t) = \frac{1}{\chi_{0,1}} 
  \int_{0}^{t} d\tau\, K_{n}(t-\tau)\, a_{0}(\tau) + \nu(t),  
\end{equation}
with the memory kernel $K_{n}(t-\tau)$ defined as
\begin{equation}
    K_{n} (t-\tau) =  \frac{(-i\gamma)^{n}}{(n-1)!} (t-\tau)^{n-1} 
                      e^{- \gamma (t-\tau)} ,
\end{equation}
and $\nu(t)$ denotes the noise operator which is a function of  $b_{k}^{(j,j+1)}(0)$ and $a_{j}(0)$. The memory Kernel corresponds to the probability density function of the Gamma distribution, and  it is noteworthy that the memory kernel $K_{n} (t-\tau)$ peaks at $t-\tau = (n-1)/\gamma$. Thus, by setting $\gamma = (n-1)/\Delta t$, we obtain a memory kernel which peaks at $\Delta t$.  Furthermore, the variance of $(t-\tau)$ can be shown to be $\sqrt{\braket{(t -\tau - \Delta t)^2} } = \frac{\sqrt{n}}{n-1} \Delta t$, which vanishes as $\Delta t/\sqrt{n}$ in the large $n$ limit. Crucially, the memory Kernel becomes in this limit a delta function
\begin{equation} \label{MemoryKernel}
    K_{n} (t-\tau) \rightarrow (-i)^{n} \delta(t - \tau - \Delta t),\,\,\text{as}\,\,n \rightarrow \infty,
\end{equation} 
which is the discrete-time memory kernel desired.
In the following, we work with system size $n\gg 1$, which renders \eqref{MemoryKernel} hold approximately. This approximation works as long as the width of $K_{n} (t-\tau)$ is negligible compared to the other timescale of site $0$. 

Finally, we can insert the relation between $a_{n}$ and $a_{0}$ back to the EOM of $a_{0}$, and obtain
\begin{equation}\label{a0EOM}
\begin{split}
    \frac{d}{dt} a_{0}(t)= &- \kappa  a_{0}(t) +i [H_{0}(t), a_{0}(t)] \\ &+ \eta \Theta(t-\Delta t) a_{0}(t-\Delta t) + \nu^{\prime}_{0}(t), 
\end{split}
\end{equation}
where the decay rate  $\kappa = (\chi_{n,0}^2 + 1/\chi_{1,0}^2 )\gamma /2 $ \footnote{Here $\kappa$ is defined as the decay rate of amplitude. The photon number decay rate is $2\kappa$.}, the feedback strength $\eta =  (-i)^{n+1}  \gamma  (   \chi_{n,0}/\chi_{0,1} ) $, 
  and the noise operator $\nu^{\prime}_{0}(t)$ is a function of $b_{k}^{(j,j+1)}(0)$ and $a_{j>0}(0)$. 
%(explicit expression in \cite{SupMat}).  

Note that \eqref{a0EOM} is the same as that in conventionally studied models with time-discrete feedback, such as the atom-mirror case (see e.g.\,\cite{GrimsmoPRL2015,PichlerPRL2016}), except the noise operator $\nu^{\prime}_{0}(t)$. Notice that $\nu^{\prime}_{0}(t)$ is a function of annihilation operators $b_{k}^{(j,j+1)}(0)$ and $a_{j>0}(0)$. When we start from vacuum initial state for the bath and the auxiliary oscillator $ \ket{\psi_{1,\ldots,n,b}(t=0)} =\ket{0}_{1} \ldots \ket{0}_{n} \ket{0}_{b}$, we have $\nu^{\prime}_{0}(t)\ket{\psi_{1,\ldots,n,b}(t=0)} =0$. So, when we consider expectation values of normally-ordered operators like $\braket{a_{0}^{\dagger}(t_1) a_{0}(t_2)}$, their EOMs necessarily become independent of $\nu^{\prime}_{0}(t)$ under this initial condition \footnote{Note that there is no requirement about initial condition of site $0$ because $\nu_{0}^{\prime}(t)$ is not a function of $a_{0}(0)$ operator. The vacuum initial state of the bath has already been assumed in deriving the zero-temperature Lindblad master equation.}. Similar conclusion holds for the atom-mirror case. That is, although these two cases have different noise operators, they give the same EOMs for the normally-ordered operators. Therefore, when considering the normally-ordered operators, we have provided a Markovian embedding of the paradigmatic atom-mirror case \footnote{Note that this is different from simply considering the total Hamiltonian of the atom and bath, which is a closed system. Here being an open system can provide computational advantage as shown later.}. The discrepancy in noise makes a difference when anti-normally-ordered operators are considered, which is left for future research.

In the following, we apply this Markovian embedding approach to a linear (a harmonic oscillator) and a nonlinear system (a qubit) with time-delayed feedback. The purpose is to give a better understanding of our approach and to show that it is a useful technique for dealing with non-Markovian dynamics efficiently.

{\it Linear case: single photon decay.---}To give an exemplary illustration of our approach, we apply it to the simplest case with a linear oscillator at site $0$. Here we study decaying dynamics of a single photon at site $0$ i.e.\,the initial state is $ \ket{\Psi_{0,1,\ldots,n}(t=0)} =\ket{1}_{0} \ket{0}_{1} \ldots \ket{0}_{n}$.  By letting the oscillator $0$ be on-resonant with the auxiliary oscillators, we have $H_{0}=0$. Since there is only one excitation in the system, this case is the same as decay of a qubit in its excited state.   

\begin{figure}[bt]
	\center
	\includegraphics[width=0.235\textwidth]{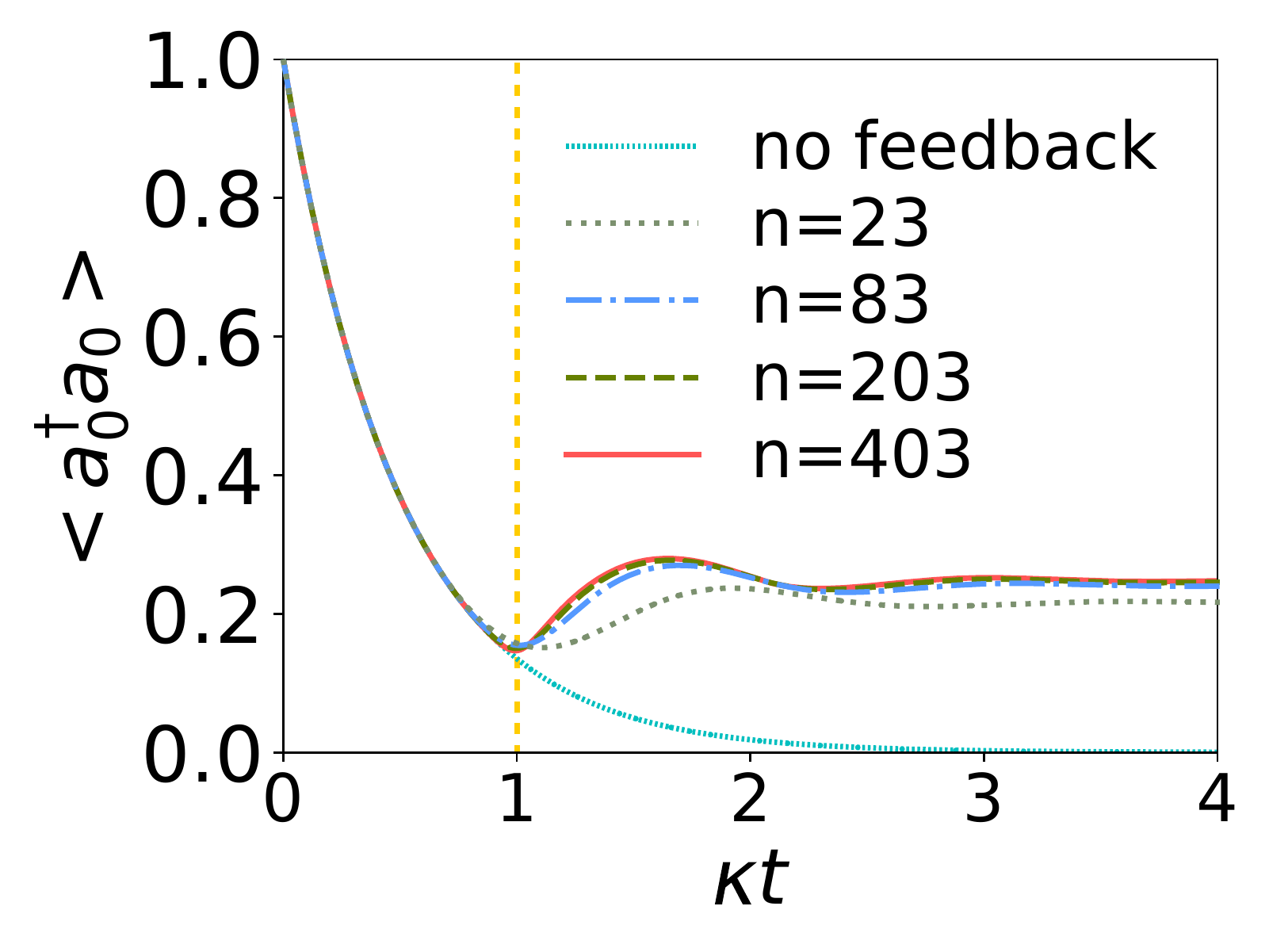}
	\includegraphics[width=0.235\textwidth]{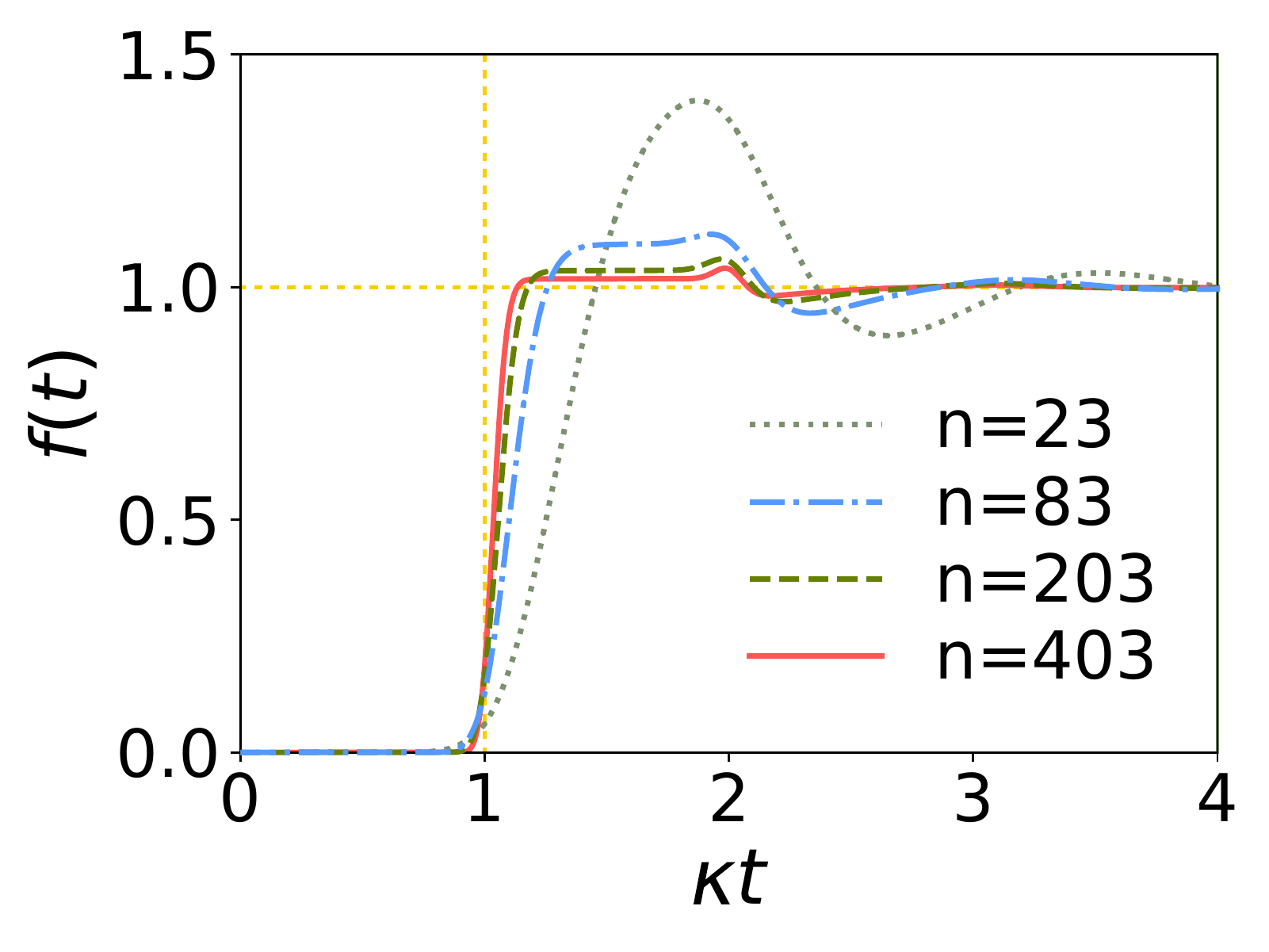}
	\put(-143,23){(a)}
    \put(-23,71){(b)}
	
	\includegraphics[width=0.235\textwidth]{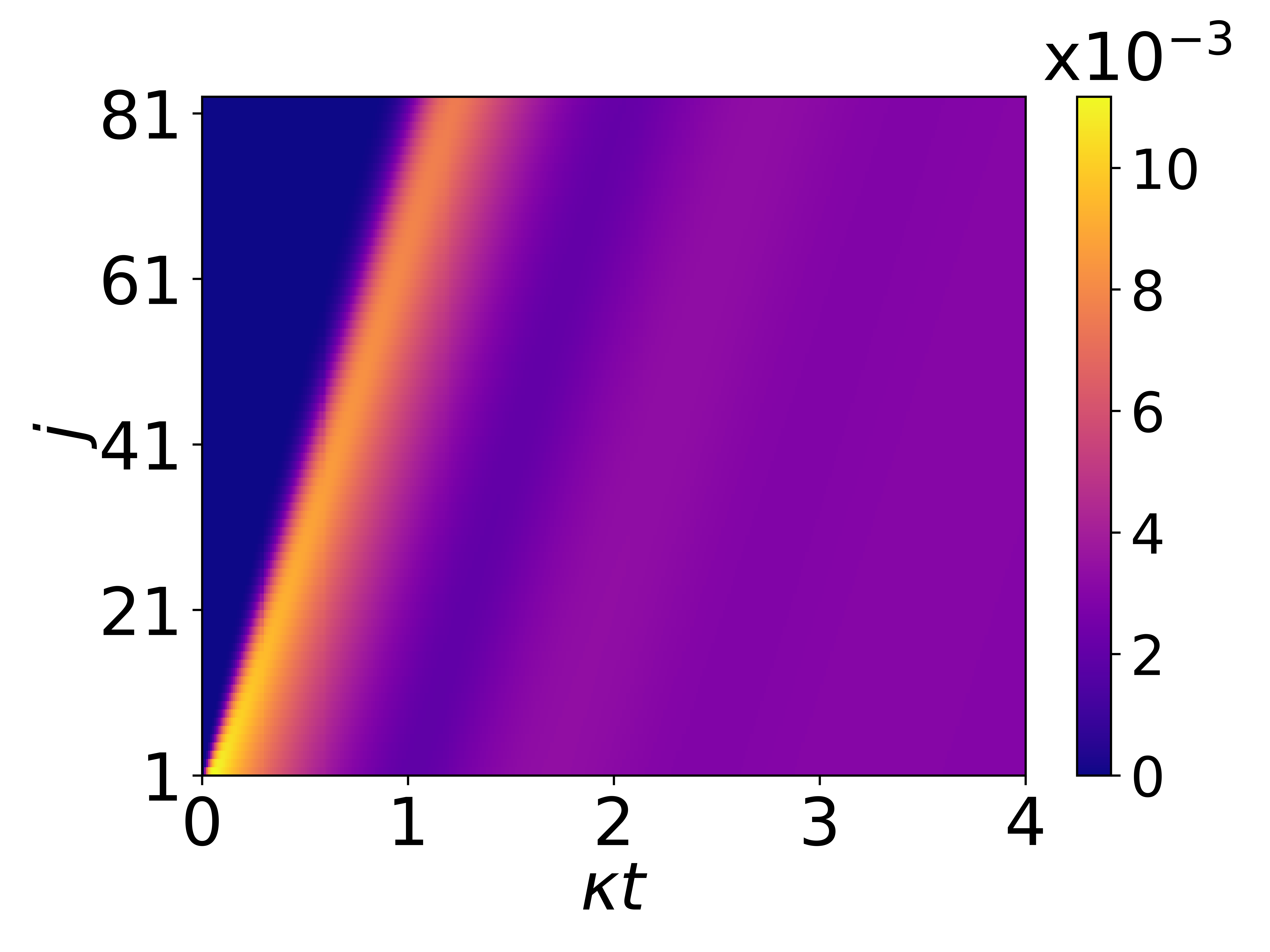}
	\includegraphics[width=0.235\textwidth]{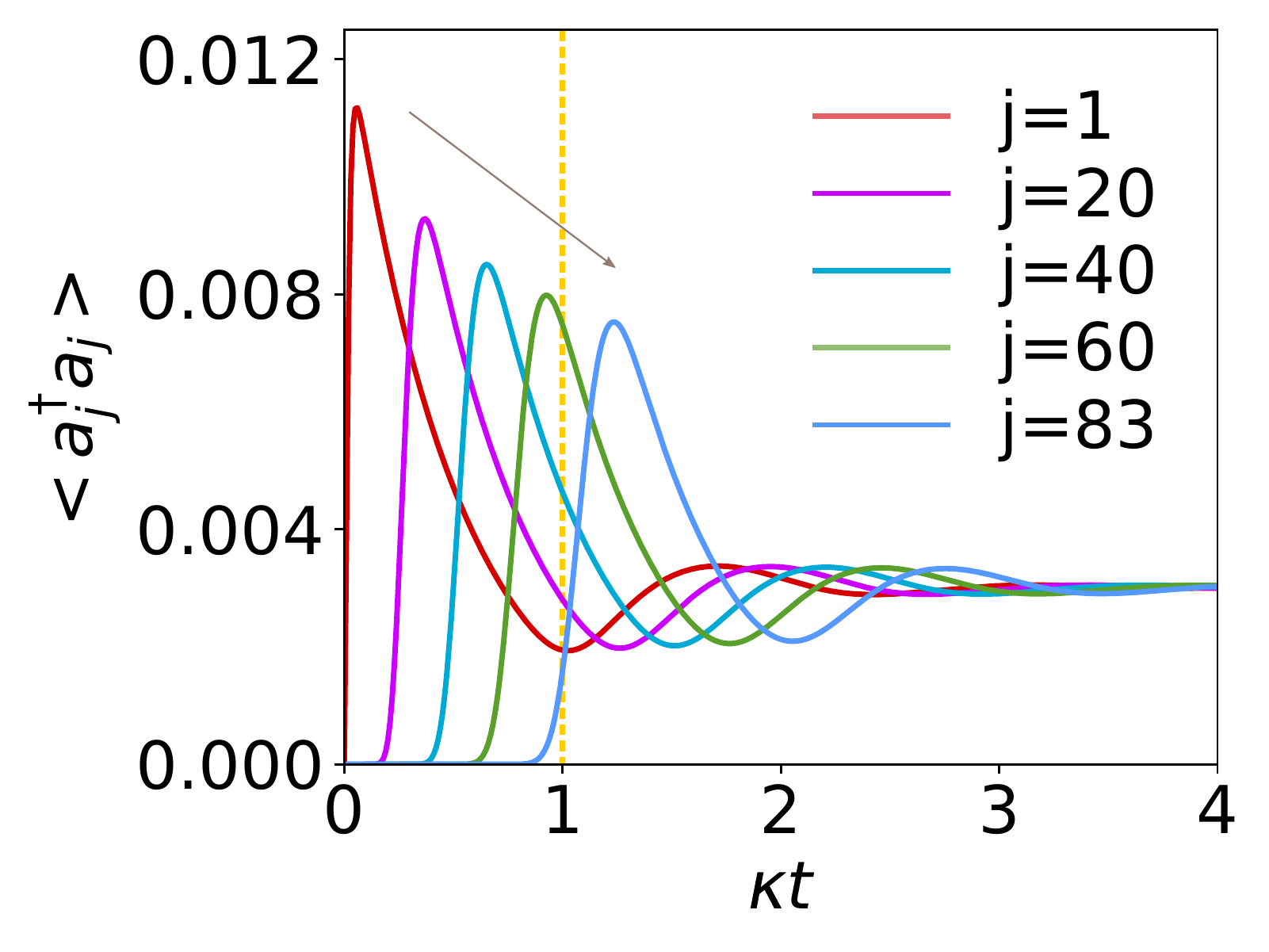}
	\put(-160,71){(c)}
    \put(-23,23){(d)}
	\caption{ Decay of a single photon with time-delayed feedback. (a) Photon population at site $0$ for $n$ auxiliary oscillators ($n=23, 83, 203, 403$). (b) The memory profile function for different size $n$. It converges to the step function in \eqref{Memoryan} as $n$ increases. (c) Population of the auxiliary oscillators in the case with $n=83$. (d) Slices of panel (c) at site $j=1, 20 , 40, 60, 83$. (values of $j$ refer curves from left to right). (Parameters: $\eta=\kappa$, $\kappa=1$, $\Delta t =1$.)  
	}
	\label{FigLinearCase}
\end{figure}

In the time-delayed EOM \eqref{a0EOM}, decay rate $\kappa$ and the feedback strength $\eta$ can be tuned with parameters  $\{ \chi_{0,1}, \chi_{n,0}, \gamma \}$, which is a minimal set of parameters we found to keep $\kappa$ and $\eta$ of order $1/\Delta t$ even in presence of the large decay rate $\gamma \sim n/\Delta t$. Here, without loss of generality, we set $\eta = \kappa$. The phase of $\eta$ can be set to be $1$ by letting $n=4m+3$ with $m\in \mathbb{Z_{+}}$. Note that phases $\{\pm,\pm i \}$ can be tuned by simply varying $n$. In general, $\eta$ can have arbitrary phase by making $\chi_{i,j+1}$ complex. Then by using the expression of $\eta$, we get our parameters in terms of $\kappa$ and $\Delta t$:
\begin{equation}
\begin{split}\label{ParameterChoice}
     \chi_{n,0} =   \sqrt{\frac{\kappa}{\gamma}},   
     \,\, \chi_{0,1} =  \sqrt{\frac{\gamma}{\kappa} },  
       \,\, 
    \gamma &= \frac{(n-1)}{\Delta t}.
\end{split}
\end{equation}

Since the EOMs \eqref{ajEOM} are linear and time-local in this case, we can calculate $\braket{a_{0}^{\dagger}(t) a_{0}(t)}$ by writing out the EOMs of $\braket{a_{j}^{\dagger}(t) a_{j^{\prime}}(t)}$. Then the matrix EOMs of these $(n+1)^2$ variables can be solved with exact diagonalization. The results are shown in Fig.\,\ref{FigLinearCase} for $\kappa=1$ and $\Delta t =1$. Fig.\,\ref{FigLinearCase}(a) shows time evolution of the $0$-th site's photon population $\braket{a_{0}^{\dagger}(t) a_{0}(t)}$ for system size $n=23, 83, 203, 403$. It can be seen that results converges quickly in $n$, and even a fairly small size $n=83$ gives a good approximation to the time-discrete feedback as predicted by our theory above. It is also easy to see the residual photon occupation that does not decay, which is exactly the well-known photon bound state in atom-mirror case (see e.g.\,\cite{GrimsmoPRL2015,PichlerPRL2016}). Here, the residual photon occupation can be understood as a result of self-driving due to feedback, which gives a nonequilibrium steady state with non-zero occupation.

To see how good a finite-sized system approximates the time-discrete feedback, we define a memory profile function $f(t) \propto  \braket{a_{n}^{\dagger}(t) a_{n}(t)}/\braket{a_{0}^{\dagger}(t-\Delta t) a_{0}(t- \Delta t)}$ \footnote{$\braket{a_{0}^{\dagger}(t-\Delta t) a_{0}(t- \Delta t)}$ is set to be $1$ for $t< \Delta t$.} according to \eqref{a0EOM}. With proper normalization, we know that $f(t) \rightarrow \Theta(t-\Delta t)$ as $n\rightarrow \infty$. Indeed, as shown in Fig.\,\ref{FigLinearCase}(b), $f(t)$ converges to the step function as $n$ increases. 

To understand the roles played by the auxiliary oscillators, their populations are shown in Fig.\,\ref{FigLinearCase}(c) and (d) for the case with $n=83$. We can see that the photon travels unidirectionally from site $1$ to site $n$ with an almost linear speed. At time $t=\Delta t$, the photon reaches the last site $n$, which turns on the time-delayed feedback. As $n$ gets larger, the turning-on process gets sharper. Finite size $n$ gives to a finite bandwidth in the auxiliary chain and therefore determines how sharp is the turning-on process for $\braket{a_{n}^{\dagger} a_{n}}$, which is the origin of non-zero width in the memory kernel $K_{n-2} (t-\tau)$.

{\it Nonlinear case: a coherently driven qubit.---}As a showcase of applying our approach to nonlinear systems, here we study a driven-dissipative qubit at site $0$. Then $H_{0}$ in \eqref{a0EOM} 
has the form $H_{0} = (\Omega a_{0}^{\dagger} + \text{h.c.}) + U a^{\dagger}_{0}a^{\dagger}_{0} a_{0}a_{0}$ with $U\rightarrow \infty$, where $\Omega$ is the driving strength and $U$ is the nonlinearity to ensure site $0$ acts as a qubit. This is equivalent to replacing the bosonic operator $a_{0}$ by the Pauli operator $\sigma^{-}_{0}$ in the master equation, i.e.\,treating $a_{0}$ as hard-core boson. Then by going through the process from the master equation to \eqref{a0EOM}, we see that, compared with bosonic representation, an additional factor $[\sigma^{+}_{0},\sigma^{-}_{0}]$ is present in the EOM of $\sigma^{-}_{0}$ \cite{SupMat}. which corresponds exactly to the EOM of a qubit in front of a mirror (see e.g.\,\cite{GrimsmoPRL2015,PichlerPRL2016}).

\begin{figure}[bt]
	\center
	\includegraphics[width=0.47\textwidth]{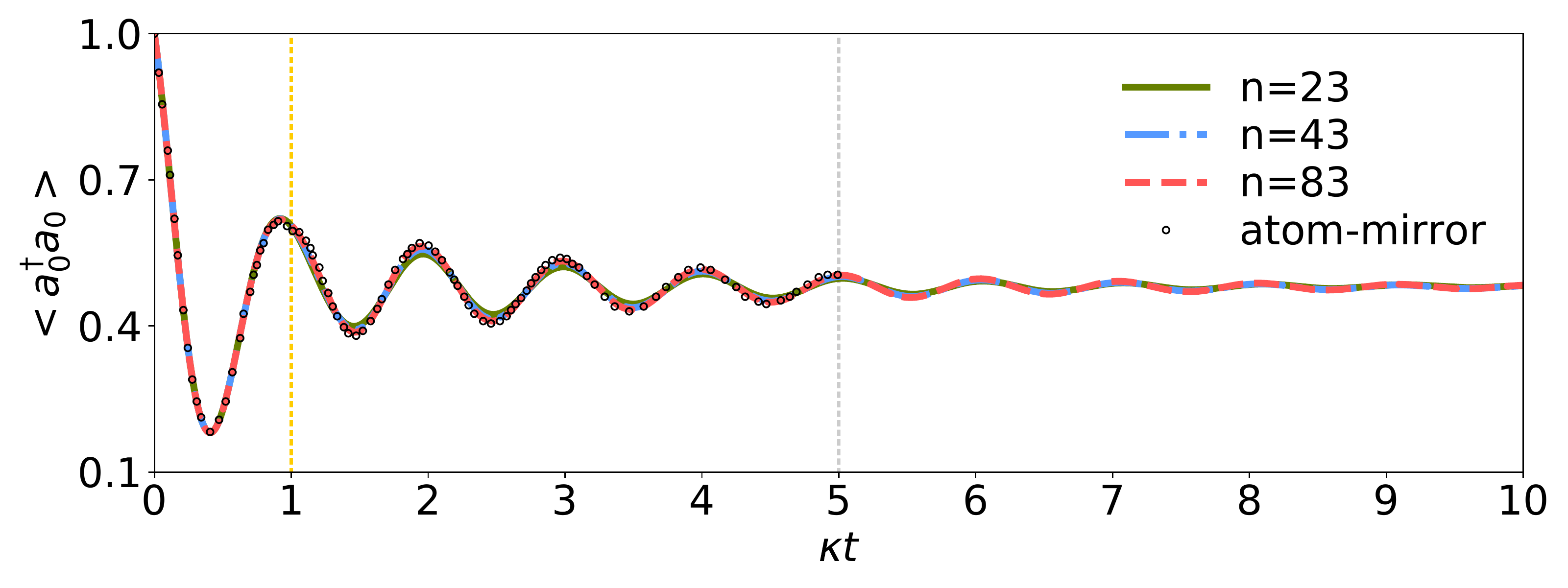}
	\put(-23,28){(a)}
	
	\includegraphics[width=0.235\textwidth]{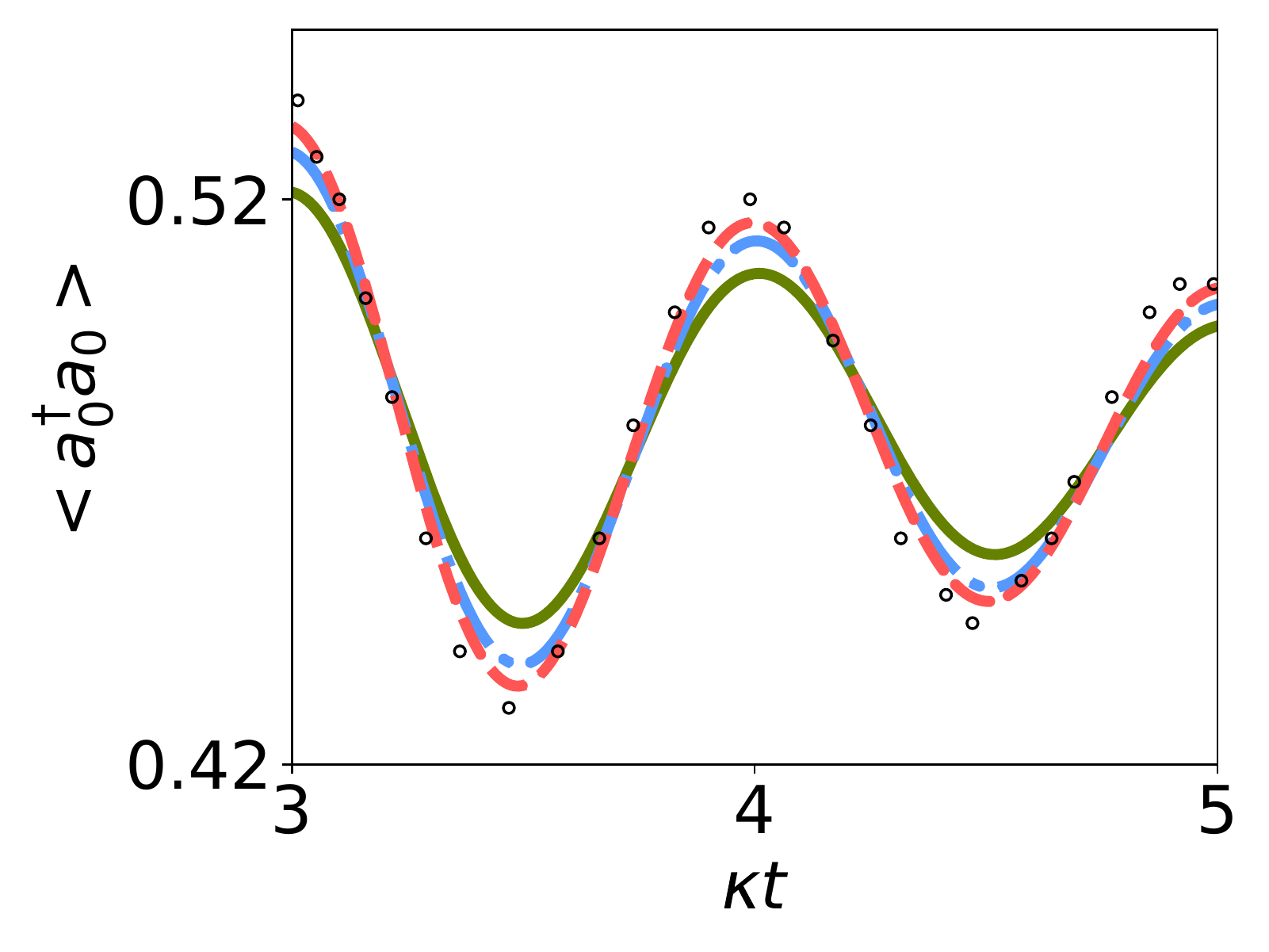}
	\includegraphics[width=0.235\textwidth]{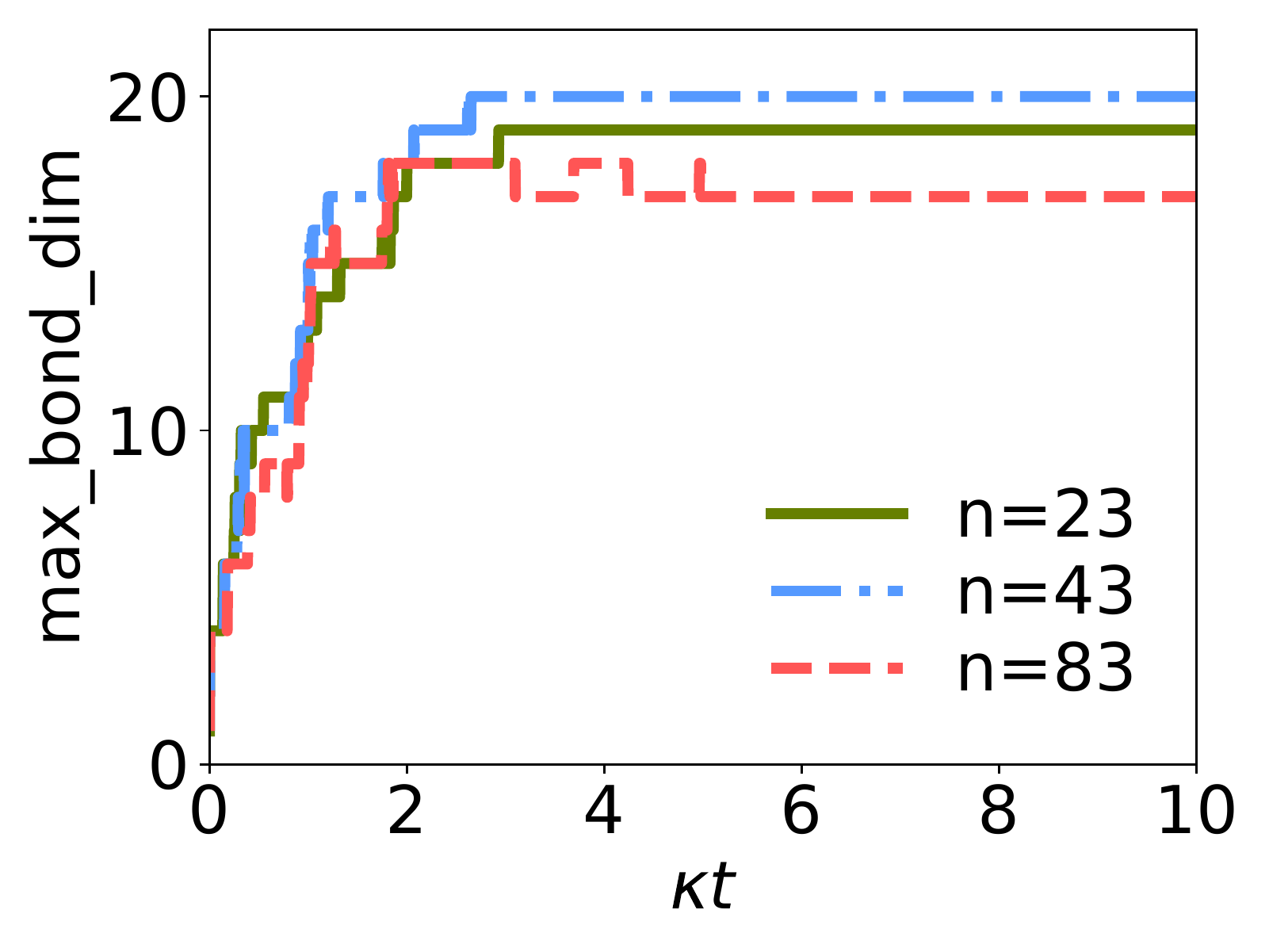}
	\put(-143,73){(b)}
    \put(-23,57){(c)}
	\caption{(a) Time evolution of qubit excited state population at site $0$ for $n=23, 43, 83$. The atom-mirror data from $t=0$ to $5$ are extracted from Fig.\,4(b) of Ref.\,\cite{GrimsmoPRL2015}. (b) Zoom-in view of panel (a) from $t=3$ to $5$. (c) Time evolution of the maximum bond dimension of the MPDO for a fixed cutoff in our MPS simulation, which serves as an indicator of computational cost.  (Parameters: $\eta=\kappa$, $\kappa=1$, $\Delta t =1$, $\Omega =\pi$; time step $\delta t =2\times 10^{-3}$, cutoff $\epsilon=10^{-10}$.)  
	}
	\label{qubit}
\end{figure}

The many-body Lindblad master equation corresponding to \eqref{a0EOM} can be solved numerically using matrix product state (MPS) methods. Here we write the density matrix as a matrix product density operator (MPDO), and then perform time evolution by applying Trotter gates onto it (see e.g.\,\cite{SCHOLLWOCKAoP2011}). It can be a problem for MPS simulation if the local Hilbert space dimension required by the auxiliary oscillators is large. Fortunately, the auxiliary oscillators here can be treated as qubits as shown in the following argument. We can write down an approximate relation between $a_{j}$ and $a_{0}$:
\begin{equation}
    a_{j}(t) \propto  \Theta(t-\Delta t_{j}) a_{0}(t-\Delta t_{j}) +  \text{noise},
\end{equation}
where $\Delta t_{j} \approx (j/n) \Delta t$. Then we know $\braket{a^{\dagger}_{j}(t) a_{j}(t)} \propto \Theta(t-\Delta t_{j}) \braket{a^{\dagger}_{0}(t-\Delta t_{j}) a_{0}(t-\Delta t_{j})}$, while $\braket{[a^{\dagger}_{j}(t)]^{q} [a_{j}(t)]^{q}} \propto \Theta(t-\Delta t_{j}) \braket{[a^{\dagger}_{0}(t-\Delta t_{j})]^{q} [a_{0}(t-\Delta t_{j})]^{q}} = 0$ for $q\geq 2$ since $a_{0}$ here represents hard-core boson. That is, in each site $j$, only the two lowest level $\{ \ket{0}, \ket{1} \}$ can be occupied. Therefore, the bosonic operators $a_{j}$ in the master equation can be effectively replaced with qubit operators $\sigma^{-}_{j}$, which greatly simplifies the MPS simulation.

For each time step $\delta t$, we decompose the superoperator $\exp{(\mathcal{L} \delta t)}$, where $\mathcal{L}$ is a Lindbladian, using a fourth-order Trotter–Suzuki decomposition (see \cite{SupMat}). Since we need $\gamma \delta t < 1$, the time step $\delta t <1/\gamma  \sim \Delta t/n$. For large $n$, this can require a large amount of Trotter gates and thus a small cutoff $\epsilon$, which limits the tractable system size. However, as shown below, a moderate-size system can be sufficient to give good approximating results. Then the numerical computation is performed using the package  \texttt{ITensors.jl} \cite{itensor}. For benchmarking purpose, we use the same parameters as those in Ref.\,\cite{GrimsmoPRL2015}, with $\Delta t=1$ and $\Omega =\pi$ such that the period of Rabi oscillation ($\pi/\Omega=1$) matches the delay time. The qubit is initially in the excited state. 

The results are shown in Fig.\,\ref{qubit}. We use $n=23$ , $43$, and $83$. Fig.\,\ref{qubit}(a) shows time evolution of the excited state population. We compare with the transient-regime data given in Ref.\,\cite{GrimsmoPRL2015}. There is remarkable agreement, even for a small system size $n=23$. As shown in Fig.\,\ref{qubit}(b), their discrepancies decreases as $n$ increases. Because the total system is an open system, the bond dimension required in MPS simulation stops growing after a while (similar to that in Ref.\,\cite{PichlerPRL2016}), as shown in Fig.\,\ref{qubit}(c). It is in stark contrast to the generally exponential increase of bond dimension in time evolution of closed systems (see e.g.\,\cite{SCHOLLWOCKAoP2011}). Thus, the computational cost in our approach is basically linear in time in long-time evolution.  This enables us to evolve the system for much longer times, even to approach its steady state. Comparison with other methods like the MPS simulation of stochastic Schr\"odinger equation \cite{PichlerPRL2016} needs to be explored in future study, especially when time delay or driving strength gets larger. Another noteworthy thing, in both Fig.\,\ref{FigLinearCase}(a) and Fig.\,\ref{qubit}(a), is that the steady state seems to be insensitive to finite size effect, which can be an advantage in future study of steady state properties.

{\it  Conclusion and outlook.---}In summary, we show how to perform Markovian embedding of time-delayed quantum feedback, using a chain of nonreciprocally coupled dissipative oscillators (see Fig.\,\ref{setup}). The original non-Markovian system, which is notoriously difficult to analyze, is thus turned into a Markovian open quantum system, which can be tackled efficiently using many-body methods. As an example, we apply this approach to the paradigmatic case with a qubit in front of a mirror, using MPS simulation. The complexity is found to be linear in time for long-time evolution, which enables us to go beyond transient regime and even to reach the nonequilibrium steady state.

 For future research, vast interesting applications of our approach await to be explored. For example, using the tunability of auxiliary oscillators, the memory kernel can be tuned for quantum control purpose. Besides the currently used reservoir engineering approach, it would be interesting to explore Markovian embedding using other approaches to nonreciprocity \footnote{see Ref.\,\cite{ClerkArxiv2022} and references therein. It is worth noting that, time-delayed feedback is found in a recent study \cite{LorenzoSciRep2021} of single photon decaying into a chiral ring environment, whose nonreciprocity is given by an effective magnetic field.}. Though a qubit is used as an example in the current study, our approach is general, with no requirement on the local Hamiltonian $H_{0}$ at site 0. Therefore, feedback on various other models, such as nonlinear oscillators or even many-body models, can be studied, using readily available methods for open quantum many-body systems \cite{WeimerRMP2021}. From a more general perspective, we have put forward an approach originally created to treat non-Markovian effects in classical stochastic systems, including feedback-controlled and active systems (see e.g.\,\cite{LoosNJP2020,DabelowPRX2019}), towards the quantum regime, which could lead to more interdisciplinary research between these fields.

\begin{acknowledgments}
Numerical data for figures are available at \href{https://doi.org/10.5281/zenodo.6380671}{https://doi.org/10.5281/zenodo.6380671}. This work is supported by the Deutsche Forschungsgemeinschaft through the Emmy Noether program (Grant No.\,ME 4863/1-1) and the project 163436311-SFB 910.
\end{acknowledgments}

%%%%%%%%%%%%%%%%%%%%%%%%%%%%%%%%%%%%%%%%%%%%%%%%
\bibliography{TimeDelay,BibFootnotes}
%%%%%%%%%%%%%%%%%%%%%%%%%%%%%%%%%%%%%%%%%%%%%%%%

%%%%%%%%%%%%%%%%%%%%%%%%%%%%%%%%%%%%%%%%%%%%%%%%%%%%%%%%%%%%%%%%%%%%%%%%
%%%%%%----------- Supplementary Materials------------------------%%%%
%%%%%%%%%%%%%%%%%%%%%%%%%%%%%%%%%%%%%%%%%%%%%%%%%%%%%%%%%%%%%%%%%%%%%%%%

\newpage

\widetext

\clearpage

\setcounter{equation}{0}
\setcounter{figure}{0}
\setcounter{table}{0}
\setcounter{page}{1}
\makeatletter
\renewcommand{\theequation}{S\arabic{equation}}
\renewcommand{\thefigure}{S\arabic{figure}}

\begin{center}
	
	{\large\bf Supplemental Material for ``Embedding Time-Delayed Quantum Feedback in a Nonreciprocal Array''}
	
	\vspace{0.5cm}
	
	Xin  H.  H.  Zhang,$^{1}$ S.  H.  L.  Klapp,$^{2}$ and  A.  Metelmann$^{1,3,4}$
	
	{\it $^{1}$Dahlem Center for Complex Quantum Systems and Fachbereich Physik, Freie Universit\"at Berlin, 14195 Berlin, German \\
	$^{2}$Institute for Theoretical Physics, Technische Universit\"at Berlin, 10623 Berlin, Germany\\
	$^{3}$Institute for Theory of Condensed Matter, Karlsruhe Institute of Technology, 76131 Karlsruhe, Germany\\
	$^{4}$Institute for Quantum Materials and Technology, Karlsruhe Institute of Technology, 76344 Eggenstein-Leopoldshafen, Germany
	}
\end{center}

In this Supplemental Material, we present  
            (i) the Markovian embedding for classical Langevin equation, 
            (ii) Heisenberg-Langevin equation for two oscillators,
            (iii) Heisenberg-Langevin equation for a chain of oscillators,
            (iv) derivation of the memory kernel for $a_{0}$,
            (v) matrix EOMs in the linear case,
            (vi) qubit EOM in the nonlinear case,
            and (vii) more information on MPS simulation of the qubit case.

%%%%%%%%%%%%%%%%%%%%%%%

\subsection{(i) the Markovian embedding for classical Langevin equation}

Here, we give a short pedagogical summary of the Markovian embedding for classical Langevin equation (for detailed studies, see e.g.\,\cite{LoosJSP2019,LoosNJP2020,CushingBook1977, SmithBook2010}).

Let's have the variable $x_{0}$ of interest, which has a non-Markovian dynamics i.e.\,a memory kernel in the equation of motion. Then let $x_{j}, j = 1, 2, \ldots n $ be the auxiliary variables that can act as a memory. The couplings are nonreciprocal: $0 \rightarrow 1 \rightarrow 2 \rightarrow \ldots \rightarrow n \rightarrow 0$.   

Starting from the Langevin equation of variable $x_{0}$, we can have 
\begin{equation}\label{x0equation}
    \lambda_{0} \frac{d}{dt} x_{0} = v_{0,0} x_{0} + f(x_{0}) + \eta_{0} + w_{0,n} x_{n},
\end{equation}
where $v_{0,0} < 0$ is a linear force, $f(x_{0})$ is a higher-order force, $\eta_{0}$ is a Gaussian white noise, and $w_{0,n}$ denotes coupling with $x_{n}$. For the linear chain acting as a memory, we have
\begin{equation}
    \lambda_{j} \frac{d}{dt} x_{j} = v_{j,j} x_{j} + \eta_{j} + w_{j,j-1} x_{j-1},
\end{equation}
where the last term represents the nonreciprocal interaction from $j-1$ to $j$.

For simplicity, we can assume the linear chain is homogeneous i.e.\,$v_{j,j}=v$, $w_{j,j-1}=w$, and $\lambda_{j}=\lambda$. Then we have ($j>0$)
\begin{equation}\label{NRx}
    \lambda \frac{d}{dt} x_{j} = v x_{j} + \eta_{j} + w x_{j-1}.
\end{equation}
Using Laplace transformation, we know
\begin{equation}
    \tilde{x}_{j}(s) = \frac{w}{\lambda s - v} \tilde{x}_{j-1}(s) + \frac{1}{\lambda s - v} \tilde{\eta}_{j}(s),
\end{equation}
where $x_{j}(0)=0$ is assumed.
Iteratively, we get
\begin{equation}\label{xnsolution}
    \tilde{x}_{n}(s) =  \left( \frac{w}{\lambda s - v} \right)^{n} \tilde{x}_{0}(s) + \sum_{k=1}^{n} \left( \frac{w}{\lambda s - v} \right)^{n-k} \frac{1}{\lambda s - v} \tilde{\eta}_{k}(s).
\end{equation}

We know the inverse Laplace transformation $\mathcal{L}^{-1} (\frac{1}{(s + \alpha)^{n}}) = \frac{1}{(n-1)!} t^{n-1} e^{-\alpha t} \Theta(t)$, where $\Theta(t)$ is a step function. We also know $\mathcal{L}^{-1} \big( f(s)g(s) \big) = \int_{0}^{t} d\tau f(\tau) g(t-\tau) $. Then from the first term of \eqref{xnsolution}, we get the memory kernel  
\begin{equation}
    K_{n} (t-\tau) = (\frac{w}{\lambda})^{n} \frac{1}{(n-1)!} (t-\tau)^{n-1} e^{\frac{v}{\lambda} (t-\tau)} 
\end{equation}
and the memory kernel for the noise $\eta_{k}$
\begin{equation}
    K_{\eta_{k}} (t-\tau)  = \frac{1}{w} K_{n-k+1} (t-\tau).  
\end{equation}

We can see that $K_{n}(t-
\tau)$ peaks at $(t-
\tau)=(n-1)\lambda/(-v)$. To get delay time $\Delta t$, we need to choose
\begin{equation}
    v =  - \frac{(n-1)\lambda}{\Delta t}.
\end{equation}
To make the memory kernel a distribution, we require normalization $ \int_{0}^{\infty} d\tau K_{n} (\tau) = 1$. That is,
\begin{equation}
    w = \frac{(n-1)\lambda}{\Delta t} = -v.
\end{equation}
In summary, we now have 
\begin{equation}
    K_{n} (t-\tau) = (\frac{n-1}{\Delta t})^{n} \frac{1}{(n-1)!} (t-\tau)^{n-1} e^{-\frac{n-1}{\Delta t} (t-\tau)} 
\end{equation}
It centers at $\Delta t$ and the variance is given by
\begin{equation}
    \sqrt{\braket{(t-\tau - \Delta t)^2} } = \frac{\sqrt{n+1}}{n-1} \Delta t \sim \frac{1}{\sqrt{n}} \Delta t,
\end{equation}
which means a sharp peak for large $n$. That is, $K_{n}(t-\tau) \approx \delta (t-\tau-\Delta t)$ for $n \gg 1$. 

Finally, \eqref{xnsolution} gives
\begin{equation}\label{xnsolutiont}
\begin{split}
     x_{n} (t) &= \int_{0}^{t} d\tau K_{n}(t-\tau) x_{0}(\tau) +  \nu(t),\\
     & \approx x_{0} (t-\Delta t)  +  \nu(t),
\end{split}
\end{equation}
where the noise $\nu(t)$ from the memory chain is given by 
\begin{equation}
    \nu(t) = \frac{\Delta t}{(n-1)\lambda}  \sum_{k=1}^{n} \int_{0}^{t} d\tau \eta_{k}(\tau) K_{n-k+1} (t-\tau). 
\end{equation}
Putting \eqref{xnsolutiont} back to \eqref{x0equation}, we get the time-delayed Langevin equation 
\begin{equation} \label{ClassicalDelayedLangevin}
    \lambda_{0} \frac{d}{dt} x_{0} = v_{0,0} x_{0} + f(x_{0}) + \eta_{0} + w_{0,n} x_{0} (t-\Delta t) + w_{0,n} \nu(t),
\end{equation}

%%%%%%%%%%%%%%%%%%%%%%%%%%
\subsection{(ii) Heisenberg-Langevin equation for two oscillators}

The dynamics of two oscillators is given by the Hamiltonian
\begin{equation}
    H_{12}= \frac{\gamma}{2} (a_{1} a_{2}^{\dagger} + h.c.)
\end{equation}
and the jump operator 
\begin{equation}
    J_{12} = \sqrt{\gamma} (a_{1} - i a_{2} ).
\end{equation}

Effectively, we can have a microscopic Hamiltonian for this Lindbladian, with an auxiliary strongly dissipative oscillator coupled with a bosonic bath:
\begin{equation}\label{microH}
    H = H_{12} + \frac{\sqrt{\kappa}}{2} ( J_{12} c^{\dagger} + \text{h.c.} ) + \int dk \,k b_{k}^{\dagger}b_{k} + \sqrt{\frac{\kappa}{2\pi}} \int dk (b_{k}c^{\dagger} + \text{h.c.} ),
\end{equation}
where $c$ is the operator for the strongly dissipative oscillator, and $b_{k}$ are the operators for the modes of the bosonic bath. For $1$, $2$, and c,  we then have a master equation 
\begin{equation}
    \frac{d}{dt} \rho_{12,c} = -i [  H_{12} + \frac{\sqrt{\kappa}}{2}( J_{12} c^{\dagger} + \text{h.c.} ), \rho_{12,c}] + \kappa \mathcal{D}[c] \rho_{12,c},
\end{equation}
which gives our master equation for $\rho_{12}$ in the large $\kappa$ limit (see \cite{GardinerZollerBook,ScullyZubairy1997,MetelmannPRX2015}):
\begin{equation}
     \frac{d}{dt} \rho_{12} = -i [  H_{12} , \rho_{12}] +  \mathcal{D}[J_{12}] \rho_{12}.
\end{equation}

Using the microscopic Hamiltonian \eqref{microH}, we have the Heisenberg EOM
\begin{equation}
    \begin{split}
        \frac{d}{dt} a_{1} &= -i \frac{\gamma}{2} a_{2} -i \frac{\sqrt{\kappa\gamma}}{2} c,\\
        \frac{d}{dt} a_{2} &=-i \frac{\gamma}{2} a_{1} + \frac{\sqrt{\kappa\gamma}}{2} c, \\
        \frac{d}{dt} c &= -i \frac{\sqrt{\kappa}}{2} J_{12} - \frac{\kappa}{2} c + F(t),\\
    \end{split}
\end{equation}
where $F(t) = -i \sqrt{\frac{\kappa}{2\pi}} \int dk b_{k}(0) e^{-ikt}$ is a noise operator and $\braket{F(t)F(t^{\prime})^{\dagger}} = \kappa \delta(t-t^{\prime})$ while other correlations vanish (i.e.\,zero-temperature bath). By formally integrating out the last one (with $\kappa \rightarrow \infty$), we get the Heisenberg-Langevin equation for $a_{1}$ and $a_{2}$:
\begin{equation}
    \begin{split}
        \frac{d}{dt} a_{1} &= - \frac{\gamma}{2} a_{1} + \eta(t),\\
        \frac{d}{dt} a_{2} &=- \frac{\gamma}{2} a_{2} -i\gamma a_{1} + i\eta(t),
    \end{split}
\end{equation}
where $\eta(t) = -i \sqrt{\gamma/\kappa} F(t)$ is a Gaussian white noise with $\braket{\eta(t)\eta^{\dagger}(t^{\prime})} = \gamma \delta(t-t^{\prime})$. Notice that the noise operator $\eta(t)$ is a function of annihilation operators $b_{k}(0)$.

%%%%%%%%%%%%%%%

\subsection{(iii) Heisenberg-Langevin equation for a chain of oscillators}

The dynamics for a chain of oscillators can be described with Lindblad master equation 
\begin{equation}\label{totalLindbladian}
\begin{split}
    \frac{d}{dt} \rho =   &-i \Big[ H_{0} + \sum_{j=0}^{n} H_{j,j+1}, \rho \Big]  \\ &+ \sum_{j=0}^{n} \Big( J_{j,j+1} \rho  J_{j,j+1}^{\dagger} - \frac{1}{2} \{ \rho ,  J_{j,j+1}^{\dagger} J_{j,j+1} \} \Big),
\end{split}
\end{equation}
where $H_{0}$ is a local Hamiltonian at site $0$, hopping Hamiltonian 
\begin{equation}
    H_{j,j+1} = \frac{\gamma_{j,j+1}}{2} (\chi_{j,j+1}^{*} a_{j} a_{j+1}^{\dagger} + \text{h.c.}),
\end{equation}
and the engineered jump operator 
\begin{equation}
    J_{j,j+1} = \sqrt{\gamma_{j,j+1}} (a_{j} - i \chi_{j,j+1} a_{j+1}).
\end{equation}
Note that although $H_{j,j+1}$ are linear, $H_{0}$ can be nonlinear.

Thus, for a chain of oscillators, we have the Heisenberg-Langevin equation
\begin{equation}\label{NRa}
    \begin{split}
        \frac{d}{dt} a_{j} &=- \frac{\gamma_{j-1,j}}{2} a_{j} - \frac{\gamma_{j,j+1}}{2} a_{j} -i\gamma_{j-1,j} a_{j-1} + i\eta_{j-1,j}(t) + \eta_{j,j+1}(t) \\
        &= -  \frac{\gamma_{j-1,j}+\gamma_{j,j+1}}{2}  a_{j} -i\gamma_{j-1,j} a_{j-1} + \xi_{j}(t),
    \end{split}
\end{equation}
where we have denoted noise operator \begin{equation}\label{defXi}
    \xi_{j}(t) = i\eta_{j-1,j}(t) + \eta_{j,j+1}(t).
\end{equation} 
We then have the following correlation relations (others vanish)
\begin{equation}
    \begin{split}
        \braket{\xi_{j}(t) \xi_{j}^{\dagger}(t^{\prime})} &= (\gamma_{j-1,j} + \gamma_{j,j+1}) \delta(t-t^{\prime}), \\
        \braket{\xi_{j}(t) \xi_{j-1}^{\dagger}(t^{\prime})} &= i \gamma_{j-1,j} \delta(t-t^{\prime}), \\
        \braket{\xi_{j-1}(t) \xi_{j}^{\dagger}(t^{\prime})} &= -i \gamma_{j-1,j} \delta(t-t^{\prime}). \\
    \end{split}
\end{equation}

%%%%%%%%%%%%%%%

\subsection{(iv) derivation of the memory kernel for $a_{0}$}

Using the fact that $\gamma = (n-3)/\Delta t$ being a large decay rate, we get $a_{1}(t)  \approx -2 i  \chi_{01} \gamma_{01} a_{0}(t) /(\chi_{01}^2\gamma_{01}+\gamma)   +  \text{noise}$ and $a_{n} (t)  \approx -2 i \gamma a_{n-1}(t)  /(\gamma_{n0}+\gamma)   +  \text{noise}$. Combining these relations with \eqref{ajs}, we get 
\begin{equation} \label{Memoryan}
    a_{n} (t)  \approx   \frac{(-i)^{n} 4 \gamma  \chi_{01} \gamma_{01} }{(\chi_{01}^2\gamma_{01}+\gamma)(\gamma_{n0}+\gamma)}  \Theta(t-\Delta t) a_{0}(t-\Delta t) +  \text{noise}.
\end{equation}

From EOM of $a_{1}$ and $a_{n}$ in \eqref{ajEOM}, we can get
\begin{equation}
\begin{split} \label{a1andan}
    a_{1} (t) &= -i \chi_{01} \gamma_{01} \int_{0}^{t} d\tau e^{-\frac{(\chi_{01}^2\gamma_{01}+\gamma)}{2}(t-\tau)}  a_{0}(\tau) +  \int_{0}^{t} d\tau e^{-\frac{(\chi_{01}^2\gamma_{01}+\gamma)}{2}(t-\tau)} \xi_{1}^{\prime}(\tau) \\
    a_{n} (t) &= -i \gamma \int_{0}^{t} d\tau e^{-\frac{(\gamma_{n0}+\gamma)}{2}(t-\tau)}  a_{n-1}(\tau) +  \int_{0}^{t} d\tau e^{-\frac{(\gamma_{n0}+\gamma)}{2}(t-\tau)} \xi_{n}^{\prime}(\tau),
\end{split}
\end{equation}
where the noise operator
\begin{equation}\label{noisejprime}
    \xi^{\prime}_{j}(t)= \xi_{j}(t) + a_j(0) \delta(t)
\end{equation}
comes from inverse Laplace transformation of  $\tilde{\xi}^{\prime}_{j}(s) = \tilde{\xi}_{j}(s) + a_j(0)$. 
Note that the exponential decay inside integral has a large decay rate $\sim \gamma/2$, which is much larger than the typical rate in site 0. We can therefore apply Markov approximation here and have the relations 
\begin{equation}
\begin{split}\label{a1a0ananm1}
    a_{1}(t) & \approx -i \chi_{01} \gamma_{01}  \frac{2}{(\chi_{01}^2\gamma_{01}+\gamma)}  a_{0}(t) + \frac{2}{(\chi_{01}^2\gamma_{01}+\gamma)} \xi_{1}^{\prime}(t)  \\
    a_{n} (t) & \approx -i \gamma   \frac{2}{(\gamma_{n0}+\gamma)}  a_{n-1}(t) +  \frac{2}{(\gamma_{n0}+\gamma)}  \xi_{n}^{\prime}(t),
\end{split}
\end{equation}
which complements the relation between $a_{n-1}$ and $a_{1}$ to give us the relation between $a_{n}$ and $a_{0}$ in \eqref{Memoryan}.

It is also straightforward to write down the expression of the noise operator $\nu^{\prime}(t)$. From \eqref{anm1s}, we know
\begin{equation}
\nu(t) = \frac{i}{\gamma} \sum_{j=2}^{n-1} \int d\tau K_{n-j} (t-\tau) \xi_{j}^{\prime} (\tau).
\end{equation}
Then using \eqref{a1a0ananm1}, we see
\begin{equation}
    \nu^{\prime} (t) \propto \sum_{j=1}^{n} \int d\tau M_{j}(t-\tau) \xi_{j}^{\prime} (\tau) + \xi_{0} (t),
\end{equation}
where $M_{j}(t-\tau)$ is their corresponding memory kernel. Notice that $\xi_{0} (t)$ only depends on $b_{k}^{(0,1)}(0)$ and $b_{k}^{(n,0)}(0)$ unlike \eqref{noisejprime}. Now we see that $\nu^{\prime} (t)$ is a function of $b_{k}^{(j,j+1)}(0)$ and $a_{j>0}(0)$.  
%%%%%%%%%%%%%%%

Interestingly, in the specific case with $\eta=\kappa$, the parameter choice \eqref{ParameterChoice} happens to give exact results without the above Markov approximation. In this case, \eqref{ParameterChoice} gives \eqref{MarkovajEOMlinear} below. Then, we get
\begin{equation}
\begin{split}
    \tilde{a}_{1}(s) &= \frac{-i\sqrt{\kappa \gamma}}{ s +\gamma} \tilde{a}_{0}(s) + \frac{1}{ s +\gamma} \left(\tilde{\xi}_{1}(s) + a_1(0)\right),\\
     \tilde{a}_{j, 2\leq j\leq n}(s) &= \frac{-i\gamma}{ s +\gamma} \tilde{a}_{j-1}(s) + \frac{1}{ s +\gamma} \left(\tilde{\xi}_{j}(s) + a_j(0)\right).
\end{split}
\end{equation}
Following the same procedure in the main text, we get
\begin{equation}
    a_{n} (t) = \sqrt{\frac{\kappa}{\gamma}} \int_{0}^{t} K_{n} (t-\tau) a_{0}(\tau)  + \text{ noise}.
\end{equation} 
Then the memory term is exactly
\begin{equation}
    -i\sqrt{\kappa \gamma} a_{n} (t) = - i \kappa \int_{0}^{t} K_{n} (t-\tau) a_{0}(\tau)  + \text{ noise},
\end{equation}
which is consistent with the general case \eqref{Memoryan}, since  $ K_{n} \rightarrow (-i)^2 K_{n-2} $ for large $n$.
%%%%%%%%%%%%%%%

\subsection{(v) matrix EOMs in the linear case}

With the parameters chosen in the linear case, \eqref{ajEOM} has the compact form
\begin{equation}\label{MarkovajEOMlinear}
\begin{split}
    \frac{d}{dt} a_{0}&= -  \kappa  a_{0} -i \sqrt{\kappa\gamma} a_{n} + \xi_{0}(t) \\   
    \frac{d}{dt} a_{1}&= -  \gamma   a_{1} -i\sqrt{\kappa \gamma}  a_{0} + \xi_{1}(t) \\ 
    \frac{d}{dt} a_{j, 2\leq j\leq n}&= -  \gamma  a_{j} -i\gamma a_{j-1} + \xi_{j}(t). \\ 
\end{split}
\end{equation}

Denote \eqref{MarkovajEOMlinear} as
\begin{equation}
    \frac{d}{dt} \Vec{a} = \mathcal{M} \Vec{a} + \Vec{\xi},
\end{equation}
with $\Vec{a} = (a_{0}, a_{1}, \ldots, a_{n})$. To get the population, we then we can solve the EOM
\begin{equation}
  \frac{d}{dt} \braket{\Vec{a^{\dagger}}\otimes \Vec{a}} = (\mathcal{M}^{*} \otimes 1 + 1 \otimes \mathcal{M} ) \braket{\Vec{a^{\dagger}}\otimes \Vec{a}},
\end{equation}
using exact diagonalization.

%%%%%%%%%%%%%%%
\subsection{(vi) qubit EOM in the nonlinear case}

In the qubit case, we have $a_{0}$ ($=\sigma_{-}$) being a hard core boson operator. 
For now, consider the bond $0\rightarrow1$ only. From \eqref{totalLindbladian}, we have
\begin{equation}
    H_{01} = \frac{\gamma_{01}\chi_{01}}{2} (a_{0} a_{1}^{\dagger} + a_{1} a_{0}^{\dagger}) 
\end{equation}
and the jump operator
\begin{equation}
    J_{01} = \sqrt{\gamma_{01}} (a_{0} - i \chi_{01} a_{1}).
\end{equation}
We can then write down the Heisenberg-Langevin equation 
\begin{equation}
\begin{split}
    \frac{d}{dt} a_{0} &= [a_{0},a_{0}^{\dagger}] ( - \frac{\gamma_{01}}{2} a_{0} + \eta_{01} ), \\
    \frac{d}{dt} a_{1} &= - i \gamma_{01}\chi_{01} a_{0} - \frac{\gamma_{01}}{2} \chi_{01}^2 a_{1} + i \chi_{01} \eta_{01}, 
\end{split}
\end{equation}
where nonlinear nature is shown as the commutation relation $[a_{0},a_{0}^{\dagger}]$ in the EOM of $a_{0}$. 

When considering other bonds, we still get $a_{n}(t) \propto \Theta(t-\Delta t) a_{0}(t-\Delta t)$ as the linear case, because EOM for $a_{n=1,\ldots,n}$ are the same. For $a_{0}$, we then have 
\begin{equation}
    \frac{d}{dt} a_{0} = [a_{0},a_{0}^{\dagger}] \big( -  \frac{1}{2}(\chi_{n0}^2\gamma_{n0}+\gamma_{01})  a_{0} -i\chi_{n0} \gamma_{n0} a_{n} + \text{ noise} \big).
\end{equation}
We then see that the term contains time-delayed feedback have the form
 $\Theta(t-\Delta t) [a_{0}(t),a_{0}^{\dagger}(t)] a_{0}(t-\Delta t)$, which is exactly the memory for qubit in front of a mirror.  Note that if $a_{0}$ is a bosonic operator, we get back to the linear case as expected.

%%%%%%%%%%%%%%%%%%%%%%%

\subsection{(vii) more information on MPS simulation of the qubit case}

Starting from the Hamiltonian and jump operator, we have the Lindbladian superoperator at bond $(j \rightarrow j+1)$
\begin{equation}
    \mathcal{L}_{j,j+1} \bullet =  -i [H_{j,j+1}, \bullet] + J_{j,j+1} \bullet J_{j,j+1}^{\dagger} - \frac{1}{2} \{ J_{j,j+1}^{\dagger}J_{j,j+1}, \bullet\}.
\end{equation}
The total Lindbladian for our system is then $\mathcal{L} = \sum_{j=0}^{n} \mathcal{L}_{j,j+1}$, which is simply a one-dimensional lattice model with periodic boundary condition (PBC). 

For the time evolution, we decompose each time step $\exp{(\mathcal{L} dt)}$ using the 4-th order Trotter decomposition (see e.g.\,\cite{BarthelAP2020}):
\begin{equation}
    e^{(\mathcal{L}_{\text{e}}+\mathcal{L}_{\text{o}})dt} = e^{\mathcal{L}_{\text{e}} dt_1} e^{\mathcal{L}_{\text{o}} dt_2} e^{\mathcal{L}_{\text{e}} dt_3} e^{ \mathcal{L}_{\text{o}} dt_4} e^{\mathcal{L}_{\text{e}} dt_3} e^{ \mathcal{L}_{\text{o}} dt_2} e^{ \mathcal{L}_{\text{e}} dt_1} + O(dt^5),
\end{equation}
where $dt_1=w dt/2$, $dt_2=wdt$, $dt_3=(1-w)dt/2$, $dt_4 = (1-2w)dt$ with $w=(2-2^{1/3})^{-1}$. Here $\mathcal{L}_{\text{e}}$ denotes bonds with $j$ being even while $\mathcal{L}_{\text{o}}$ denotes bonds with $j$ being odd. 

Notice that $dt_1>0$ and $dt_2>0$ make $\exp(\mathcal{L}dt_{1(2)})$ a completely positive and trace-preserving (CPTP) map, while $dt_3<0$ and $dt_4<0$ make $\exp(\mathcal{L}dt_{3(4)})$ stop being a CPTP map but just a linear trace-preserving map. That is, we can perform Kraus decomposition of  $\exp(\mathcal{L}dt_{1(2)}) \rho$:
\begin{equation}
    e^{\mathcal{L}dt_{1(2)} }\rho = \sum_{k} A_{k} \rho A_{k}^{\dagger},
\end{equation}
with $\sum_{k} A_{k}^{\dagger} A_{k} =1$. We then apply operator $A_{k}$ and $A_{k}^{\dagger}$ on the MPDO to apply a Trotter gate. As for the linear map $\exp(\mathcal{L}dt_{3(4)})$, we can have an operator-sum representation \cite{Milz2017} 
\begin{equation}
    e^{\mathcal{L}dt_{3(4)}} = \sum_{k} L_{k} \rho R_{k}^{\dagger},
\end{equation}
which is used for applying a Trotter gate. These decompositions are carried out using the {\tt to\_kraus} and {\tt to\_stinespring} functions of the QuTiP library \cite{qutip2}.

\begin{figure}[bt]
	\center
	\includegraphics[width=0.33\textwidth]{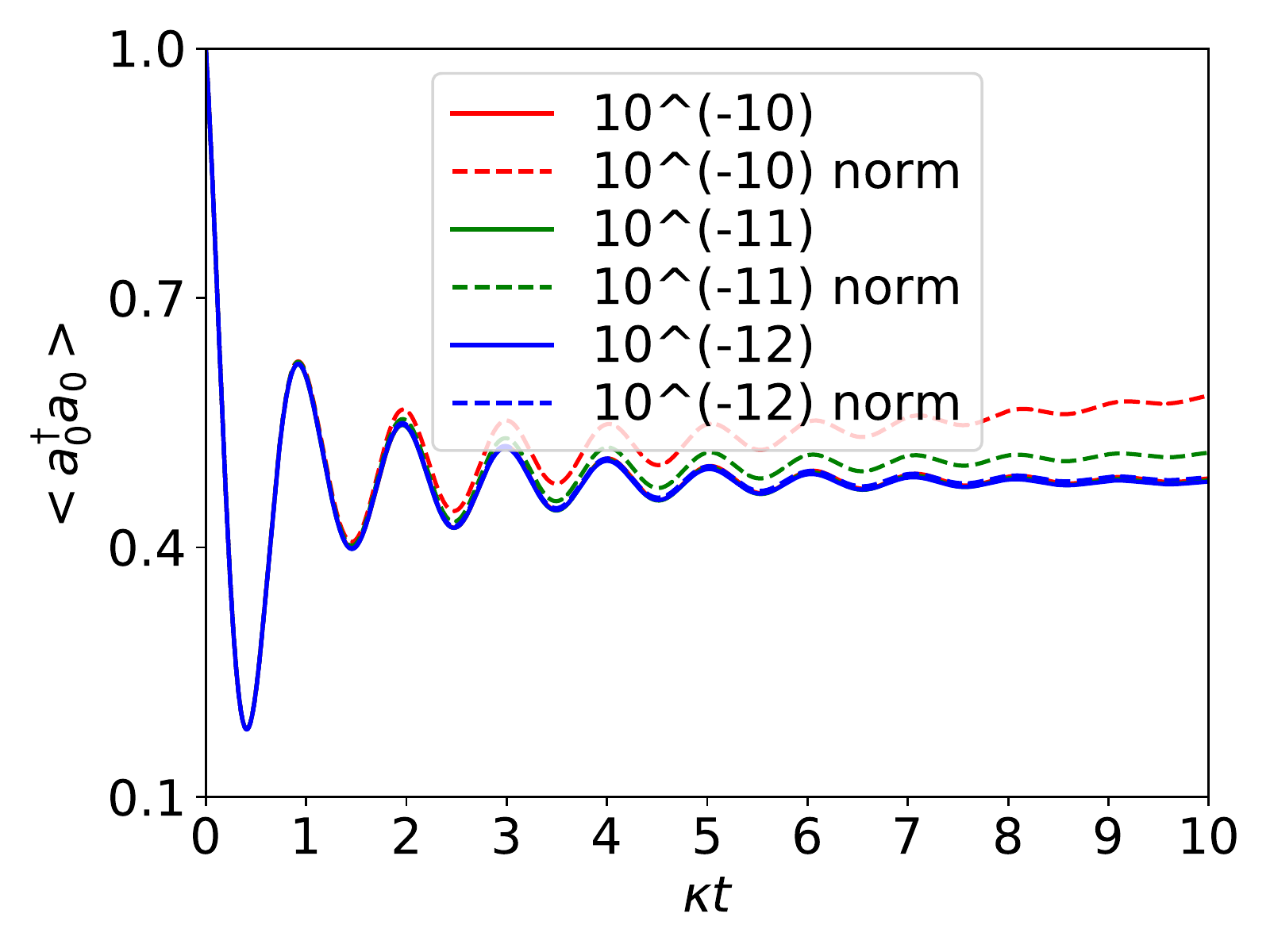}
	\put(-23,28){(a)}
	\includegraphics[width=0.33\textwidth]{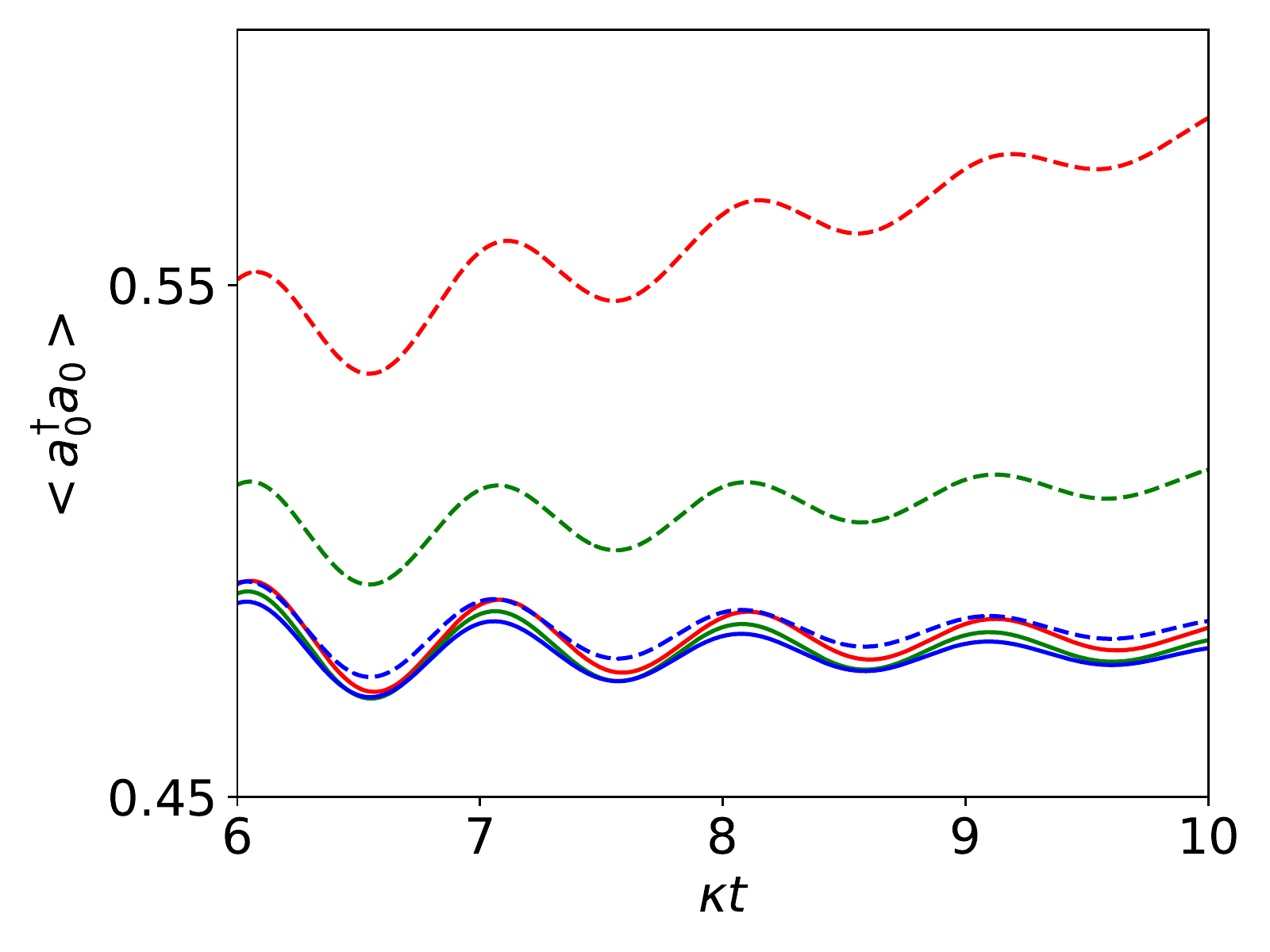}
    \put(-23,73){(b)}
	\includegraphics[width=0.33\textwidth]{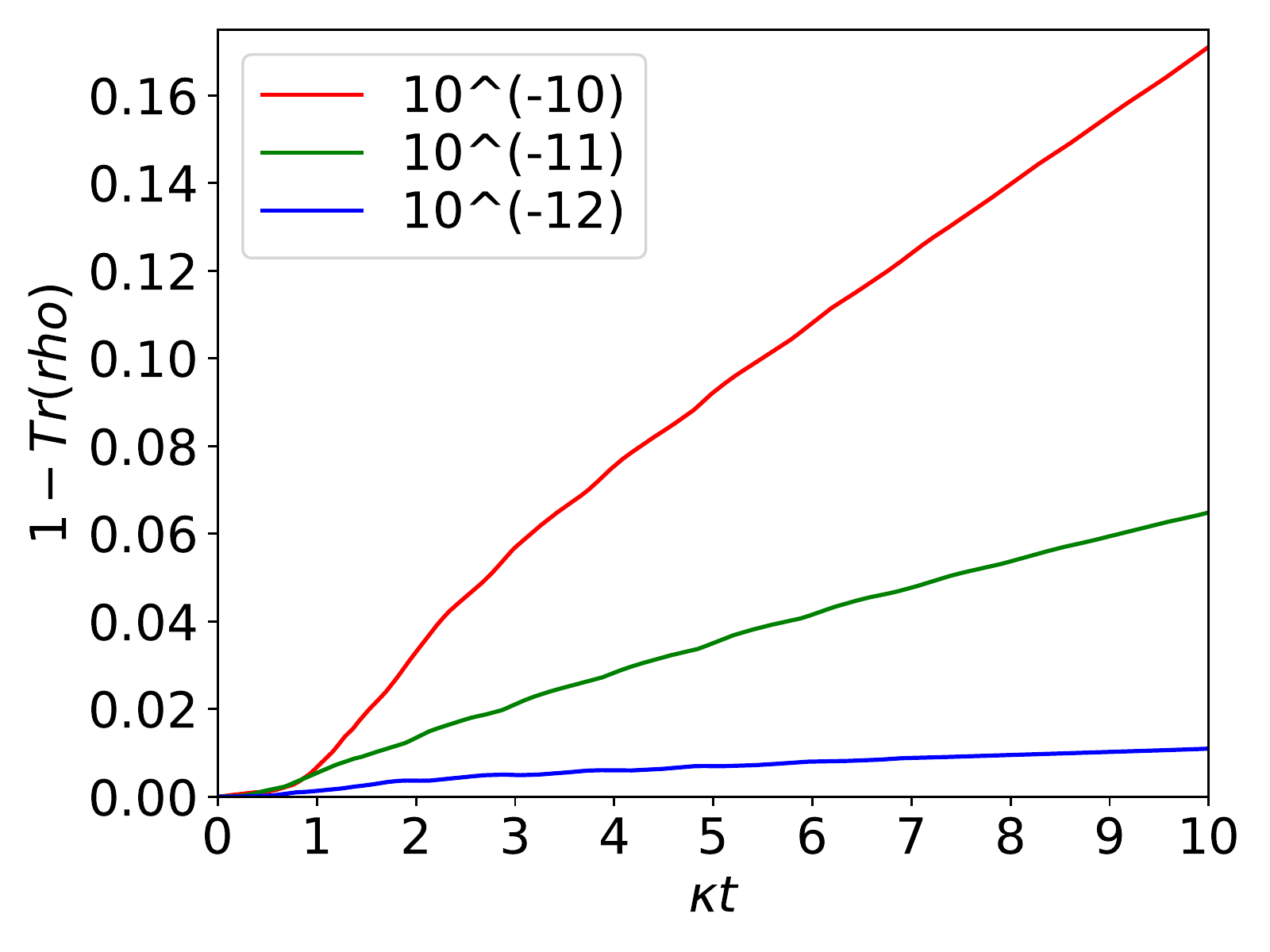}
    \put(-23,73){(c)}
	\caption{(a) Time evolution of qubit excited state population at site $0$ for $n=23$. The solid and dashed lines are for unnormalized and normalized $\rho$ respectively, with cutoff error $\epsilon=10^{-10},10^{-11}, 10^{-12} $. (b) Zoom-in view of panel (a) from $t=6$ to $10$. (c) The trace error $1-\Tr(\rho)$ during MPS time evolution.  (Parameters: $\eta=\kappa$, $\kappa=1$, $\Delta t =1$, $\Omega =\pi$, $n=23$; time step $\delta t =2\times 10^{-3}$.)  
	}
	\label{SupNorm}
\end{figure}

As mentioned in the main text, small time step means requirement for smaller cutoff error $\epsilon$ in MPS simulation. The density matrix with $\epsilon=0$ ($\equiv \tilde{\rho}$) can be written as 
\begin{equation}
    \tilde{\rho} = \rho + \mathcal{E} (\epsilon) \rho_{x},
\end{equation}
where $\rho$ is the MPDO we have in simulation, $\rho_{x}$ (already normalized) is the part that got truncated away, and $\mathcal{E} (\epsilon)$ is an unknown function of the cutoff $\epsilon$. Because $\mathcal{E} (\epsilon) $ is non-zero for finite $\epsilon$. That is, we have trace error $1 - \Tr(\rho) = \mathcal{E} (\epsilon)$ [see Fig.\,\ref{SupNorm}(c)]. In our computation, we compute $\braket{a_{0}^{\dagger}a_{0}}$ using $\rho$ instead of the normalized $\rho^{\prime} = \rho / \Tr(\rho)$ based on the following observation. As shown in Fig.\,\ref{SupNorm}(a) and (b), as $\epsilon$ gets smaller, the unnormalized values changes little compared with the normalized ones. We attribute this to the fact that, beyond a cutoff $\epsilon > 10^{-10}$ here, the truncated part $\rho_{x}$ almost does not contribute to the computed observable  $\braket{a_{0}^{\dagger}a_{0}}$. That is, $\Tr(a_{0}^{\dagger}a_{0} \rho_{x} ) \approx 0$. Based on this property, we use the unnormalized $\rho$ in our computation of $\braket{a_{0}^{\dagger}a_{0}}$. Of course, systematic error analysis will be needed in future studies.

%%%%%%%%%%%%%%%%%%%%%%%

\end{document}